\documentclass[aps,prl,reprint,superscriptaddress,showpacs]{revtex4-1} 
\usepackage{amsmath}
\usepackage{amssymb}
\usepackage{graphicx}
\usepackage{color}
\newcommand{\bsl}{\boldsymbol}

\begin{document}

\preprint{}

\title{Single Dirac point and helical states in a one-dimensional system}
\author{Sheng-Nan \surname{Ji}}
\affiliation{State Key Laboratory of Low-Dimensional Quantum Physics, and Department of Physics, Tsinghua University, Beijing, 100084, China}
\affiliation{Department of Physics and Institute of Theoretical Physics, The Chinese University of Hong Kong, Shatin, N.T., Hong Kong, China}
\author{Bang-Fen \surname{Zhu}}\thanks{bfz@mail.tsinghua.edu.cn}
\affiliation{State Key Laboratory of Low-Dimensional Quantum Physics, and Department of Physics, Tsinghua University, Beijing, 100084, China}
\affiliation{Institute of Advanced Study, Tsinghua University, Beijing 100084, China}
\author{Ren-Bao \surname{Liu}}\thanks{rbliu@phy.cuhk.edu.hk}
\affiliation{Department of Physics and Institute of Theoretical Physics, The Chinese University of Hong Kong, Shatin, N.T., Hong Kong, China}
\date{\today}

\begin{abstract}
Odd numbers of Dirac points and helical states can exist at edges (surfaces) of two-dimensional (three-dimensional) topological insulators. In the bulk of a one-dimensional lattice (not an edge) with time reversal symmetry, however, a no-go theorem forbids the existence of an odd number of Dirac points or helical states. Introducing a magnetic field can violate the time reversal condition but would usually lift the degeneracy at the Dirac points.
We find that a spatially periodic magnetic field with zero mean value can induce a single Dirac point in a one-dimensional system with spin-orbit coupling. A wealth of new physics may emerge due to the existence of a single Dirac point and helical states in the bulk of a one-dimensional lattice (rather than edge states). A series of quantized numbers emerge due to the non-trivial topology of the 1D helical states, including the doubled period of helical Bloch oscillations, quantized conductance near the Dirac point, and 1/2-charge solitons at mass kinks. Such a system can be realized in one-dimensional semiconductor systems or in optical traps of atoms.
\end{abstract}


\maketitle

Topological insulators have non-trivial topological structures in their bulk states as compared with the vacuum~\cite{TI,dirac1}. Therefore at the edges of two-dimensional (2D) or the surfaces of three-dimensional topological insulators there must be gapless states, characterized by an odd number of Dirac points~\cite{dirac1,dirac2,dirac3,dirac4,dirac5,dirac6,dirac7}. The states around the Dirac points have the spin orientation locked to the motion direction, called helical states~\cite{helical}. Exotic magneto-optical and transport phenomena are associated with the helical liquids at the edges or surfaces of topological insulators~\cite{TI,dirac1,nogo,half2,halfchelical}. In the bulk of one-dimensional~(1D) systems (not the edges of a 2D system) that have time-reversal symmetry, however, a no-go theorem~\cite{nogo} forbids the existence of an odd number of Dirac points. Introducing a magnetic field can violate the time-reversal condition of the no-go theorem, but would generally lift the degeneracy at the Dirac points and destroy the helical liquid. A pure 1D helical liquid would be the platform for a vast range of new physics~\cite{nogo,half2,halfchelical,helical1}. Therefore it is highly desirable to design a 1D system to bypass the no-go theorem.

\begin{figure}[t]
\includegraphics[width=\columnwidth]{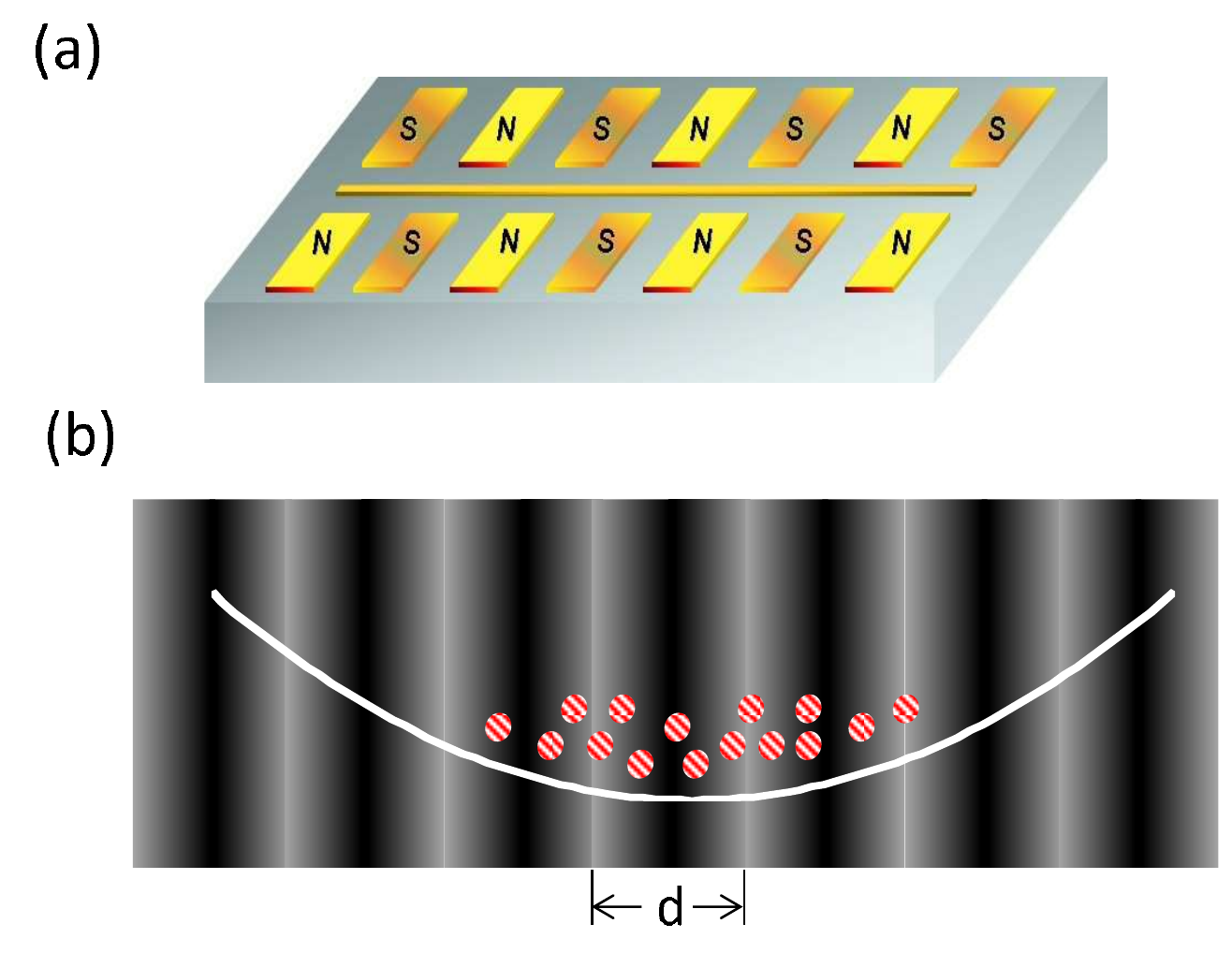}
\caption{\label{fig-system}(Color online) (a) A model system based on a semiconductor quantum wire fabricated between two periodic alignments of alternately plated magnets. (b) A model system based on a 1D optical trap (the parabola) of cold atoms with spin-orbit coupling and a periodic magnetic field (background) induced by Raman couplings.}
\end{figure}

In this letter, we report the discovery of a single Dirac point and helical states in spin-orbit coupling~(SOC) 1D systems under a spatially periodic magnetic field. A 1D system with the SOC and time-reversal symmetry has a crossing point at momentum $k=0$ between the energy bands associated with the two spin states. Then, a spatially periodic magnetic field with period $d$ opens an energy gap at quasimomentum  $k=\pm \pi/d$ [the edges of the Brillouin zone in the magnetic lattice]. The magnetic field has a zero mean value and therefore preserves the degeneracy at $k=0$, forming
a single Dirac point. The model is analogous to the Haldane model in a 2D lattice in the presence of a spatially periodic magnetic field~\cite{Haldane}. The single Dirac point and the associated helical states in the bulk of the 1D system lead to a series of exotic phenomena characterized by quantized numbers. For example, under a strong electric field $F$, the electrons would perform a helical Bloch oscillation with a period $4\pi\hbar/(eFd)$, twice the period of the conventional Bloch oscillation. Also, the conductance is quantized to be $e^2/h$ or 0 for a Fermi energy within or without the Dirac cone region, respectively. The quantized conductance is robust against slow varying disorders. Furthermore, at mass kinks, the system has soliton solutions with fractional charge $e/2$. These predicted phenomena would be just a few examples of a vast range of new physics to be explored in the 1D Dirac fermion
system.

Various physical systems may be adopted to realize the model studied in this paper. One can fabricate a semiconductor quantum wire between two alignments of alternately plated magnets [as shown in Fig.~\ref{fig-system}(a)] or in a confined magnonic crystal~\cite{magnonic}. Another possibility is to
use cold atoms in one-dimensional optical traps. Raman transitions in the atoms can be used to induce the SOC~\cite{MIT,ZhaiHui} as well as the magnetic field~\cite{olsf} [as shown in Fig.~\ref{fig-system}(b)]. To be specific, we employ the semiconductor system in Fig.~\ref{fig-system}(a) for model study.

The effective Hamiltonian of the proposed 1D model reads
\begin{eqnarray}
\label{eq-Hamiltonian}
\mathcal{H}=-\frac{\hbar^2}{2m^*}\frac{\partial^2}{\partial x^2}+i\mathbf{A}\cdot\bsl{\sigma}\frac{\partial}{\partial x}+\mathbf{B}(x)\cdot\bsl{\sigma},
\end{eqnarray}
where $\mathbf{A}$ is the SOC coefficient, $\bsl{\sigma}$ is the spin operator, $m^*$ is the effective mass of the electron, and $\mathbf{B}(x)=\mathbf{B}(x+d)$ is the periodic magnetic field with a period $d$. We take the $x$-component of the vector potential of the magnetic field to be zero, so the momentum $k$ is a good quantum number.

\begin{figure}[t]
\includegraphics[width=\columnwidth]{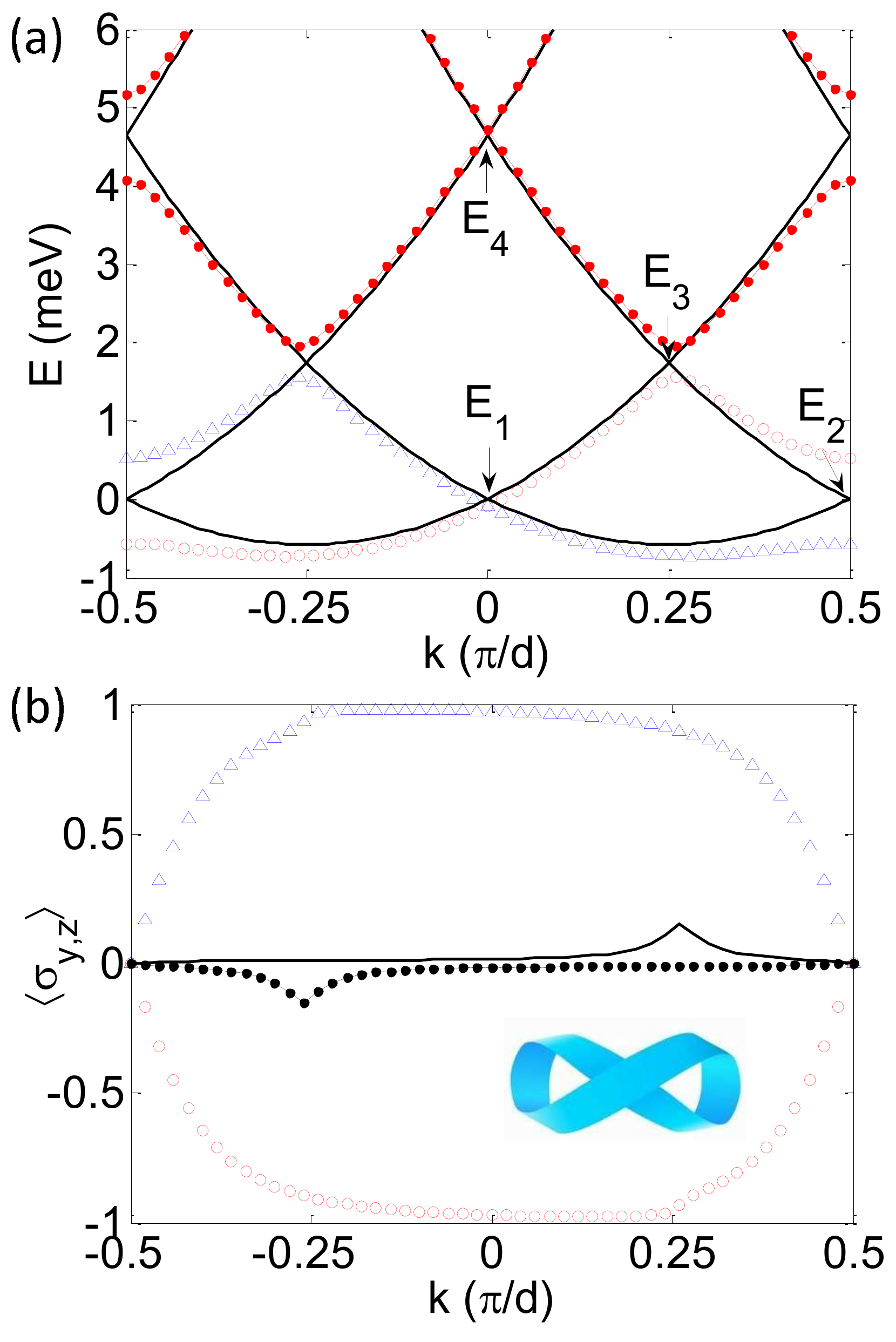}
\caption{\label{fig-energy}(Color online) (a) Solid (black) lines represent the calculated energy bands without the magnetic field; the red circles, blue triangles, and red dots correspond to the $E_{-}$-band, $E_{+}$-band and the excited bands in the presence of the periodic magnetic field.
(b) Expectation values of $\sigma_y$ and $\sigma_z$ in the two lowest energy bands. The red circles and blue triangles represent $\langle\sigma_y\rangle$ in the energy bands $E_+$ and $E_-$, respectively. The black solid line and the black dots represent $\langle\sigma_z\rangle$ in the energy band $E_+$ and $E_-$, respectively. The inset illustrates the non-trivial twisting topology of the energy bands.}
\end{figure}

Figure~\ref{fig-energy}(a) shows the formation of the single Dirac point. Let us first consider the case $\mathbf{B}(x)=0$. For a given SOC coefficient $\mathbf{A}$, the electron with momentum $k$ experiences an effective magnetic field $k\mathbf{A}$. Therefore the energy bands for the two spin states quantized along $\mathbf{A}$ are shifted in the momentum space by
\begin{eqnarray}
\label{eq-kso}
k_{so}={2m^*|\mathbf{A}|}/{\hbar^2}.
\end{eqnarray}
The two energy parabola of the two spin states cross at $k=0$ (the $E_1$ point in the figure).
A periodic magnetic field $\mathbf{B}(x)$ will fold the energy bands into the Brillouin zone $[-\pi/d, \pi/d]$. The
folded energy bands have degeneracy points at the Brillouin zone edge ($E_2$) and other points ($E_3, E_4, \ldots$).
We define the $n$th order Fourier coefficient of the magnetic field as
\begin{eqnarray}
\label{eq-field 1st F}
\mathbf{B}_n=d^{-1}\int_0^d \mathbf{B}(x)e^{-i2n\pi x/d}dx.
\end{eqnarray}
If $\mathbf{B}_n\ne 0$ and not parallel or anti-parallel to $\mathbf{A}$, the magnetic field will mix the spin states at the degeneracy points where the momentum difference (before folding) between the states is $2n\pi/d$ and open gaps between the minibands (with the gap $\sim B_n$). Since the mean value of the magnetic field is zero (i.e., $B_0=0$), the degeneracy at $k=0$ (the $E_1$ point) is preserved. If we choose the period of the magnetic field $d \sim \pi/k_{so}$, the energy band crossing at $k=0$ would be located in the gap between the minibands, becoming a single Dirac point. Thus the helical states are formed at the Dirac point.

We take a CdSe quantum wire as the 1D model system~\cite{CdSe}. The effective mass of electrons $m^*=0.22m_0$~\cite{effectmass} and the Lande g-factor $g=2$~\cite{g-factor}.  With a reasonable value of the spin-orbit coupling $A=2\times10^{-11}$~eVm (assumed along the $y$-axis), we get $k_{so}=1.16\times10^{8}$~m$^{-1}$, and the period of the magnetic field is taken as $d=27.0$~nm, which is much larger than the lattice constant of the material. Under such conditions, the Hamiltonian Eq.~(\ref{eq-Hamiltonian}) validates as an effective model near the $\Gamma$ point~($k=0$). 
For simplicity, we take the periodic magnetic field as
\begin{eqnarray}
\label{eq-Bfieldcos}
\mathbf{B}(x)=\mathbf{B}\cos\left({2\pi x}/{d}\right),
\end{eqnarray}
where $\mathbf{B}=\left(0,B_y,B_z\right)$ with $B_y=0.20$~meV and $B_z=0.55$~meV, corresponding to a magnetic field strength $5.03$~Tesla. Figure.~\ref{fig-energy}~(a) shows the calculated energy bands without the magnetic field~(black solid curves folded into the Brillouin zone) and with the magnetic field~(the curves with symbols). The two lowest minibands are separated by gaps from the high-lying bands. As expected, there is an energy crossing point at $k=0$. We denote the two lowest bands as $E_+$-band~(the red circles) with negative spin polarization along the $y$-axis ($\langle\sigma_y\rangle<0$), and the $E_-$-band~(the blue triangles) with $\langle\sigma_y\rangle>0$, respectively.  The energy gap at the Brillouin zone edge is about $1.1$~meV, nearly $2B_z$, or equivalent to about $13$~K in the scale of temperature.

To check the helical nature of the states at the Dirac point, we now analyze the spin polarizations in the two lowest energy bands. Figure~\ref{fig-energy}~(b) shows that electrons with opposite wave vectors have opposite spin polarizations. The spin polarizations are normal
to the wavevectors. In the vicinity of the Dirac point, the spin is quantized along
$\mathbf{A}$. With increasing $k$, the electron state changes gradually to a mixed state (due to spin-momentum entanglement). At the Brillouin zone edge, both $\langle\sigma_y\rangle$ and $\langle\sigma_z\rangle$  become zero so that the wave functions satisfy the periodic boundary condition as
\begin{eqnarray}
\langle\psi_{+}(\pi/d)|\psi_{-}(-\pi/d)\rangle = 1.
\end{eqnarray}
Therefore, if we connect the state $\psi_{+}(\pi/d)$ with $\psi_{-}(-\pi/d)$ and $\psi_{+}(-\pi/d)$ with $\psi_{-}(\pi/d)$, a twisted loop with a single node at the center of the Brillouin zone, like a $\infty$-shape knot, is constructed~[see the inset of Fig.~\ref{fig-energy}~(b)]. This topological structure contrasts sharply with the ordinary periodic system in which the loop is simply a circle without nodes.

\begin{figure}[t]
\includegraphics[width=\columnwidth]{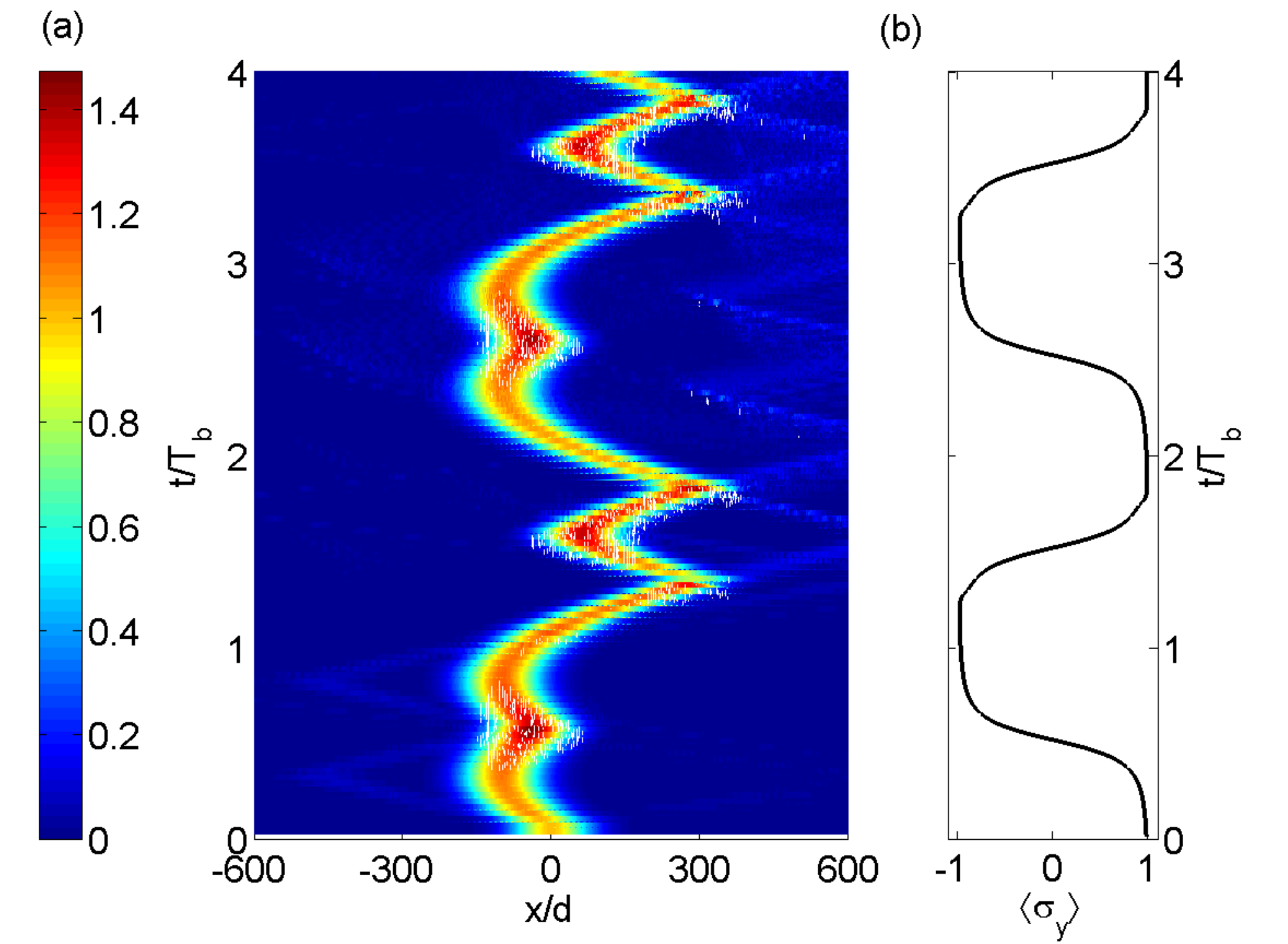}
\caption{\label{fig-3}(Color online)
Helical Bloch oscillation. (a) The absolute value of the spin-resolved electron probability amplitude $|U(x,t)|$~(unnormalized) as a function of position and time.  (b) Spin polarization $\langle\sigma_y\rangle$ of the electron as a function of time.}
\end{figure}

To study the global physical effects due to the non-trivial topology of the helical states, we consider an electron moving adiabatically along the whole energy minibands. This can be realized by applying an electric field to drive the electron across the Broullion zone. The electron would
move along one energy band (e.g., the $E_+$-band) and then at the Brillouin zone edge enter into the other energy band. It would have to across the Brillouin zone twice before completing a periodic motion. Similar to the conventional Bloch oscillation~\cite{BOsemi1,BOoptical}, the electron would oscillate periodically in real space, but with its spin state twisted periodically. We call such an oscillation the helical Bloch oscillation. The period of the helical oscillation is $4\pi\hbar/(eFd)=2T_b$, twice that of the conventional Bloch oscillation. Thus the period of the (helical) Bloch oscillation can be taken as a quantum number characterizing the topology of the 1D electrons. We note that the helical Bloch oscillation would not occur at the edges of a two-dimensional topological insulator since the gapless edge states always merge into the bulk states at higher energy.

For numerical investigations of the helical Bloch oscillation, we use the time-dependent Schr\"{o}dinger equation
\begin{eqnarray}
i\hbar\frac{\partial}{\partial t}\psi(x,t)=\left(\mathcal{H}-eFx\right)\left(\begin{array}{c}
U(x,t) \\
D(x,t)
\end{array}
 \right),
\end{eqnarray}
where $\mathcal{H}$ is the Hamiltonian in Eq.~(\ref{eq-Hamiltonian}) and $U(x,t)$ and $D(x,t)$ are wavefunctions associated with the $\sigma_z$-up and $\sigma_z$-down, respectively. The calculated absolute value of the spin-resolved probability amplitude $|U(x,t)|$ is shown in Fig.~\ref{fig-3}(a). Since the energy gap at $E_3$ is rather small, to reduce the Landau-Zener tunneling to higher bands the field cannot be too strong ($eFd\ll 1$~meV). On the other hand, the oscillation period $2T_b$ should be less than the scattering relaxation time. In our calculation we take the static electric field of $F=2.4$~kV/m (accordingly $2T_b=20$~ps).  We choose a wave-packet initially prepared in a pure $\sigma_y$-up state in the $E_-$ band with the wavevector centered at $k=0$, so $|U(x,t)|=|D(x,t)|$. Under an electric field $F$, the wave-packet moves to the left at $t=0$. After arriving at the leftmost of the oscillation (the bottom of the $E_-$-band), the electron turns to right and gradually changes its spin state. At $t=T_b/2$, the electron arrives at the rightmost of the oscillation in a spin mixed state [at the energy band $E_-(\pi/2d)$, which is equivalent to the state at $E_+(-\pi/2d)$]. After that the electron moves in the $E_+$-band, through a left minimum and a right maximum, and at $t=3T_b/2$ it is back into the $E_-$-band again. At time $t=2T_b$ the electron returns to the initial state. The light blue lines on the background in Fig.~\ref{fig-3} with small amplitudes are due to Landau-Zener tunneling to higher bands.  Accompanying the spatial oscillation, the spin polarization also oscillates periodically [Fig. 3(b)].

\begin{figure}[t]
\includegraphics[width=\columnwidth]{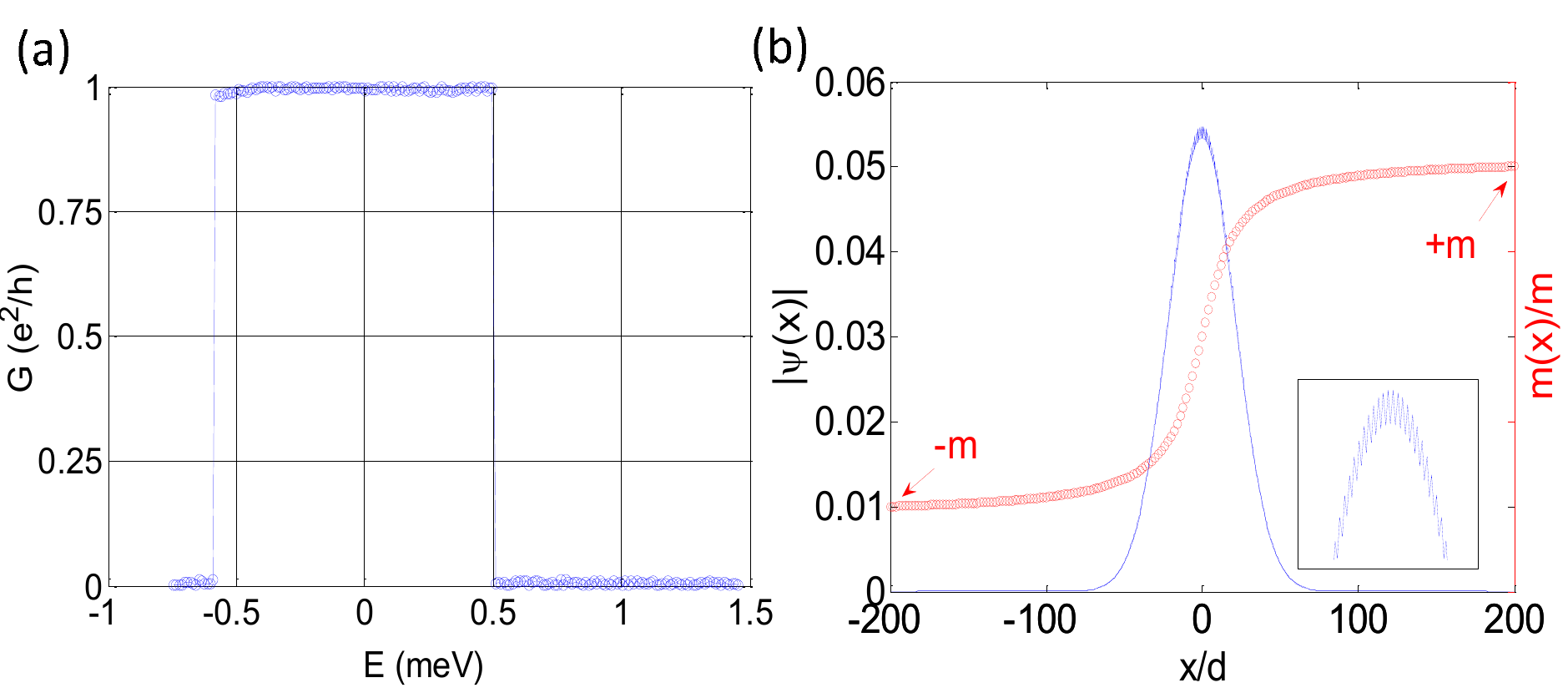}
\caption{\label{fig-2}(Color online) (a) Conductance in the 1D helical system, quantized to be $e^2/(2\pi\hbar)$ or 0 when the Fermi surface is within or without the gap containing the Dirac point, respectively. (b) Soliton solution~(solid blue curve) when there is a mass kink~(red circles) in the system. The envelop shows the wavefunction at $t=1$~ns, which is initially a normalized Gaussian wavepacket (at $t=0$). After evolution, the wavepacket has the same envelope as the initial state except for some small rapid oscillations (amplified in the inset).}
\end{figure}

The 1D Dirac electrons also have unique transport properties in the weak field regime. In particular, the helical states near the Dirac point are immune of scattering by slow-varying potential conserving time-reversal symmetry.
Similar to the configuration in Ref.~\cite{1DS2}, we consider the conductance of the 1D system subjected to $N$ random square barriers described by the potential function
\begin{eqnarray}
\label{eq-scatter}
V(x)=\left\{
\begin{array}{cc}
V_n , &  x_n\leq x\leq x_n',\\
0 ,   &  \text{others},
\end{array}
\left(n=1,2,...,N\right),\right.
\end{eqnarray}
where $x_{n+1}-x_n=l$ and $x_n'-x_n=l_{d}=0.4 l$ are fixed, and the strength $V_n$ is a uniform random number $\in [0, 1.5]$~meV.
We calculate the conductance of electrons in the two lowest bands using the Landauer-B\"{u}ttiker formula~\cite{LB}
\begin{eqnarray}
\label{eq-Landauer}
G(E)={e^2}/(2\pi\hbar)\text{Tr}(tt^\dagger),
\end{eqnarray}
where the transmission amplitude $t$ (as a function of the Fermi energy) is calculated by the transfer matrix method~\cite{GF}.
The numerical results reveal a quantized conductance of $e^2/h$ for Fermi energy around the Dirac point in spite of the sacttering [Fig.~\ref{fig-2}(a)].
For Fermi energy outside the gap that contains the Dirac point, the conductance drops to zero immediately~[Fig.~\ref{fig-2}~(a)]. Such behaviors
can be well understood using the helical state picture. Near the Dirac point, an incoming electron with a wave vector $k$ and its scattered state with $-k$ have opposite spins, so the elastic backscattering is forbidden, similar to the edge states in topological insulators~\cite{dirac3}. On the other hand, the electron with energy out of the gap region, as an electron in 1D system, is always localized in the presence of disorders~\cite{1Dlocal1}.

Another exotic phenomenon of 1D helical liquids is solitions with fractional charge. In 1976, Jackiw and Rebbi predicted that a fractional charge $e/2$ (soliton)~is carried by a mass domain wall (mass kink) in a 1D Dirac model~\cite{half1}. Later, this was proved to exist in a helical liquid system~\cite{halfchelical}. Having only half the degrees of freedom of the conventional 1D system,  the helical liquid avoids the fermion doubling problem~\cite{half2}, and therefore can carry a half charge. In the numerical calculation, we set an initial state as a normalized Gaussian wavepacket $\psi_s(x)=\mathcal{C}\exp(-x^2/\Delta x^2)|\uparrow\rangle_y$, $\mathcal{C}=(\pi\Delta x^2/2)^{-1/4}$, $\Delta x=30d$, at a mass domain of the form $m^*(x)=\arctan(0.06x/d)$ [the red circles in Fig.~\ref{fig-2}~(b)]. After a long time of evolution ($t=1$~ns), the electron has diffused but left behind a localized wavepacket with almost the same envelope as the initial state [the blue solid curve in Fig.~\ref{fig-2}(b)]. This means that there is a soliton solution in this 1D system. By integrating the charge density distribution of the localized wavepacket , we obtain
\begin{eqnarray}
\rho=e\int dx |\psi_s(x,t)|^2\approx0.5e,
\end{eqnarray}
confirming the existence of a soliton solution with fractional charge $e/2$.

In summary, we have discovered the existence of a single Dirac point and hence helical states in the bulk of a 1D model. The spin orbit coupling splits the two spin states, and a periodic magnetic field with vanishing mean value opens a gap at the Brillouin zone edges while preserving the degeneracy at the Dirac point. Exotic phenomena characterized by quantized numbers are predicted for the 1D helical states, including the helical Bloch oscillation,
quantized conductance, and soliton solutions with fractional charge. The model, realizable in 1D semiconductor systems or in 1D optical traps of cold atoms, provides a new platform for studying a wealth of physics in 1D helical liquids without requiring reduction from a higher-dimensional host system (in contrast to edge states of two-dimensional topological insulators).

\begin{acknowledgments}
This work is supported by Hong Kong RGC/GRF 401011, CUHK Focused Investments Scheme, NSFC Grant No.~11074143, and the Program of Basic Research Development of China Grant No. 2011CB921901.  We are grateful to J. Wang and C. X. Liu for useful discussions.
\end{acknowledgments}

\bibliography{1Dref}

\begin{thebibliography}{28}%
\makeatletter
\providecommand \@ifxundefined [1]{%
 \@ifx{#1\undefined}
}%
\providecommand \@ifnum [1]{%
 \ifnum #1\expandafter \@firstoftwo
 \else \expandafter \@secondoftwo
 \fi
}%
\providecommand \@ifx [1]{%
 \ifx #1\expandafter \@firstoftwo
 \else \expandafter \@secondoftwo
 \fi
}%
\providecommand \natexlab [1]{#1}%
\providecommand \enquote  [1]{``#1''}%
\providecommand \bibnamefont  [1]{#1}%
\providecommand \bibfnamefont [1]{#1}%
\providecommand \citenamefont [1]{#1}%
\providecommand \href@noop [0]{\@secondoftwo}%
\providecommand \href [0]{\begingroup \@sanitize@url \@href}%
\providecommand \@href[1]{\@@startlink{#1}\@@href}%
\providecommand \@@href[1]{\endgroup#1\@@endlink}%
\providecommand \@sanitize@url [0]{\catcode `\\12\catcode `\$12\catcode
  `\&12\catcode `\#12\catcode `\^12\catcode `\_12\catcode `\%12\relax}%
\providecommand \@@startlink[1]{}%
\providecommand \@@endlink[0]{}%
\providecommand \url  [0]{\begingroup\@sanitize@url \@url }%
\providecommand \@url [1]{\endgroup\@href {#1}{\urlprefix }}%
\providecommand \urlprefix  [0]{URL }%
\providecommand \Eprint [0]{\href }%
\providecommand \doibase [0]{http://dx.doi.org/}%
\providecommand \selectlanguage [0]{\@gobble}%
\providecommand \bibinfo  [0]{\@secondoftwo}%
\providecommand \bibfield  [0]{\@secondoftwo}%
\providecommand \translation [1]{[#1]}%
\providecommand \BibitemOpen [0]{}%
\providecommand \bibitemStop [0]{}%
\providecommand \bibitemNoStop [0]{.\EOS\space}%
\providecommand \EOS [0]{\spacefactor3000\relax}%
\providecommand \BibitemShut  [1]{\csname bibitem#1\endcsname}%
\let\auto@bib@innerbib\@empty
\bibitem [{\citenamefont {Qi}\ and\ \citenamefont {Zhang}(2011)}]{TI}%
  \BibitemOpen
  \bibfield  {author} {\bibinfo {author} {\bibfnamefont {X.-L.}\ \bibnamefont
  {Qi}}\ and\ \bibinfo {author} {\bibfnamefont {S.-C.}\ \bibnamefont {Zhang}},\
  }\href {\doibase 10.1103/RevModPhys.83.1057} {\bibfield  {journal} {\bibinfo
  {journal} {Rev. Mod. Phys.}\ }\textbf {\bibinfo {volume} {83}},\ \bibinfo
  {pages} {1057} (\bibinfo {year} {2011})}\BibitemShut {NoStop}%
\bibitem [{\citenamefont {Hasan}\ and\ \citenamefont {Kane}(2010)}]{dirac1}%
  \BibitemOpen
  \bibfield  {author} {\bibinfo {author} {\bibfnamefont {M.~Z.}\ \bibnamefont
  {Hasan}}\ and\ \bibinfo {author} {\bibfnamefont {C.~L.}\ \bibnamefont
  {Kane}},\ }\href {\doibase 10.1103/RevModPhys.82.3045} {\bibfield  {journal}
  {\bibinfo  {journal} {Rev. Mod. Phys.}\ }\textbf {\bibinfo {volume} {82}},\
  \bibinfo {pages} {3045} (\bibinfo {year} {2010})}\BibitemShut {NoStop}%
\bibitem [{\citenamefont {Bernevig}\ \emph {et~al.}(2006)\citenamefont
  {Bernevig}, \citenamefont {Hughes},\ and\ \citenamefont {Zhang}}]{dirac2}%
  \BibitemOpen
  \bibfield  {author} {\bibinfo {author} {\bibfnamefont {B.~A.}\ \bibnamefont
  {Bernevig}}, \bibinfo {author} {\bibfnamefont {T.~L.}\ \bibnamefont
  {Hughes}}, \ and\ \bibinfo {author} {\bibfnamefont {S.-C.}\ \bibnamefont
  {Zhang}},\ }\href {\doibase 10.1126/science.1133734} {\bibfield  {journal}
  {\bibinfo  {journal} {Science}\ }\textbf {\bibinfo {volume} {314}},\ \bibinfo
  {pages} {1757} (\bibinfo {year} {2006})}\BibitemShut {NoStop}%
\bibitem [{\citenamefont {K{\"o}nig}\ \emph {et~al.}(2007)\citenamefont
  {K{\"o}nig}, \citenamefont {Wiedmann}, \citenamefont {Br{\"u}ne},
  \citenamefont {Roth}, \citenamefont {Buhmann}, \citenamefont {Molenkamp},
  \citenamefont {Qi},\ and\ \citenamefont {Zhang}}]{dirac3}%
  \BibitemOpen
  \bibfield  {author} {\bibinfo {author} {\bibfnamefont {M.}~\bibnamefont
  {K{\"o}nig}}, \bibinfo {author} {\bibfnamefont {S.}~\bibnamefont {Wiedmann}},
  \bibinfo {author} {\bibfnamefont {C.}~\bibnamefont {Br{\"u}ne}}, \bibinfo
  {author} {\bibfnamefont {A.}~\bibnamefont {Roth}}, \bibinfo {author}
  {\bibfnamefont {H.}~\bibnamefont {Buhmann}}, \bibinfo {author} {\bibfnamefont
  {L.~W.}\ \bibnamefont {Molenkamp}}, \bibinfo {author} {\bibfnamefont {X.-L.}\
  \bibnamefont {Qi}}, \ and\ \bibinfo {author} {\bibfnamefont {S.-C.}\
  \bibnamefont {Zhang}},\ }\href {\doibase 10.1126/science.1148047} {\bibfield
  {journal} {\bibinfo  {journal} {Science}\ }\textbf {\bibinfo {volume}
  {318}},\ \bibinfo {pages} {766} (\bibinfo {year} {2007})}\BibitemShut
  {NoStop}%
\bibitem [{\citenamefont {Liu}\ \emph {et~al.}(2008)\citenamefont {Liu},
  \citenamefont {Hughes}, \citenamefont {Qi}, \citenamefont {Wang},\ and\
  \citenamefont {Zhang}}]{dirac4}%
  \BibitemOpen
  \bibfield  {author} {\bibinfo {author} {\bibfnamefont {C.}~\bibnamefont
  {Liu}}, \bibinfo {author} {\bibfnamefont {T.~L.}\ \bibnamefont {Hughes}},
  \bibinfo {author} {\bibfnamefont {X.-L.}\ \bibnamefont {Qi}}, \bibinfo
  {author} {\bibfnamefont {K.}~\bibnamefont {Wang}}, \ and\ \bibinfo {author}
  {\bibfnamefont {S.-C.}\ \bibnamefont {Zhang}},\ }\href {\doibase
  10.1103/PhysRevLett.100.236601} {\bibfield  {journal} {\bibinfo  {journal}
  {Phys. Rev. Lett.}\ }\textbf {\bibinfo {volume} {100}},\ \bibinfo {pages}
  {236601} (\bibinfo {year} {2008})}\BibitemShut {NoStop}%
\bibitem [{\citenamefont {Essin}\ and\ \citenamefont {Moore}(2007)}]{dirac5}%
  \BibitemOpen
  \bibfield  {author} {\bibinfo {author} {\bibfnamefont {A.~M.}\ \bibnamefont
  {Essin}}\ and\ \bibinfo {author} {\bibfnamefont {J.~E.}\ \bibnamefont
  {Moore}},\ }\href {\doibase 10.1103/PhysRevB.76.165307} {\bibfield  {journal}
  {\bibinfo  {journal} {Phys. Rev. B}\ }\textbf {\bibinfo {volume} {76}},\
  \bibinfo {pages} {165307} (\bibinfo {year} {2007})}\BibitemShut {NoStop}%
\bibitem [{\citenamefont {Hsieh}\ \emph {et~al.}(2008)\citenamefont {Hsieh},
  \citenamefont {Qian}, \citenamefont {Wray}, \citenamefont {Xia},
  \citenamefont {Hor}, \citenamefont {Cava},\ and\ \citenamefont
  {Hasan}}]{dirac6}%
  \BibitemOpen
  \bibfield  {author} {\bibinfo {author} {\bibfnamefont {D.}~\bibnamefont
  {Hsieh}}, \bibinfo {author} {\bibfnamefont {D.}~\bibnamefont {Qian}},
  \bibinfo {author} {\bibfnamefont {L.}~\bibnamefont {Wray}}, \bibinfo {author}
  {\bibfnamefont {Y.}~\bibnamefont {Xia}}, \bibinfo {author} {\bibfnamefont
  {Y.~S.}\ \bibnamefont {Hor}}, \bibinfo {author} {\bibfnamefont
  {R.}~\bibnamefont {Cava}}, \ and\ \bibinfo {author} {\bibfnamefont {M.~Z.}\
  \bibnamefont {Hasan}},\ }\href@noop {} {\bibfield  {journal} {\bibinfo
  {journal} {Nature}\ }\textbf {\bibinfo {volume} {452}},\ \bibinfo {pages}
  {970} (\bibinfo {year} {2008})}\BibitemShut {NoStop}%
\bibitem [{\citenamefont {Zhang}\ \emph {et~al.}(2009)\citenamefont {Zhang},
  \citenamefont {Liu}, \citenamefont {Qi}, \citenamefont {Dai}, \citenamefont
  {Fang},\ and\ \citenamefont {Zhang}}]{dirac7}%
  \BibitemOpen
  \bibfield  {author} {\bibinfo {author} {\bibfnamefont {H.}~\bibnamefont
  {Zhang}}, \bibinfo {author} {\bibfnamefont {C.-X.}\ \bibnamefont {Liu}},
  \bibinfo {author} {\bibfnamefont {X.-L.}\ \bibnamefont {Qi}}, \bibinfo
  {author} {\bibfnamefont {X.}~\bibnamefont {Dai}}, \bibinfo {author}
  {\bibfnamefont {Z.}~\bibnamefont {Fang}}, \ and\ \bibinfo {author}
  {\bibfnamefont {S.-C.}\ \bibnamefont {Zhang}},\ }\href@noop {} {\bibfield
  {journal} {\bibinfo  {journal} {Nat. Phys.}\ }\textbf {\bibinfo {volume}
  {5}},\ \bibinfo {pages} {438} (\bibinfo {year} {2009})}\BibitemShut {NoStop}%
\bibitem [{\citenamefont {Raghu}\ \emph {et~al.}(2010)\citenamefont {Raghu},
  \citenamefont {Chung}, \citenamefont {Qi},\ and\ \citenamefont
  {Zhang}}]{helical}%
  \BibitemOpen
  \bibfield  {author} {\bibinfo {author} {\bibfnamefont {S.}~\bibnamefont
  {Raghu}}, \bibinfo {author} {\bibfnamefont {S.~B.}\ \bibnamefont {Chung}},
  \bibinfo {author} {\bibfnamefont {X.-L.}\ \bibnamefont {Qi}}, \ and\ \bibinfo
  {author} {\bibfnamefont {S.-C.}\ \bibnamefont {Zhang}},\ }\href {\doibase
  10.1103/PhysRevLett.104.116401} {\bibfield  {journal} {\bibinfo  {journal}
  {Phys. Rev. Lett.}\ }\textbf {\bibinfo {volume} {104}},\ \bibinfo {pages}
  {116401} (\bibinfo {year} {2010})}\BibitemShut {NoStop}%
\bibitem [{\citenamefont {Wu}\ \emph {et~al.}(2006)\citenamefont {Wu},
  \citenamefont {Bernevig},\ and\ \citenamefont {Zhang}}]{nogo}%
  \BibitemOpen
  \bibfield  {author} {\bibinfo {author} {\bibfnamefont {C.}~\bibnamefont
  {Wu}}, \bibinfo {author} {\bibfnamefont {B.~A.}\ \bibnamefont {Bernevig}}, \
  and\ \bibinfo {author} {\bibfnamefont {S.-C.}\ \bibnamefont {Zhang}},\ }\href
  {\doibase 10.1103/PhysRevLett.96.106401} {\bibfield  {journal} {\bibinfo
  {journal} {Phys. Rev. Lett.}\ }\textbf {\bibinfo {volume} {96}},\ \bibinfo
  {pages} {106401} (\bibinfo {year} {2006})}\BibitemShut {NoStop}%
\bibitem [{\citenamefont {Qi}\ \emph {et~al.}(2008)\citenamefont {Qi},
  \citenamefont {Hughes},\ and\ \citenamefont {Zhang}}]{half2}%
  \BibitemOpen
  \bibfield  {author} {\bibinfo {author} {\bibfnamefont {X.-L.}\ \bibnamefont
  {Qi}}, \bibinfo {author} {\bibfnamefont {T.~L.}\ \bibnamefont {Hughes}}, \
  and\ \bibinfo {author} {\bibfnamefont {S.-C.}\ \bibnamefont {Zhang}},\
  }\href@noop {} {\bibfield  {journal} {\bibinfo  {journal} {Nat. Phys.}\
  }\textbf {\bibinfo {volume} {4}},\ \bibinfo {pages} {273} (\bibinfo {year}
  {2008})}\BibitemShut {NoStop}%
\bibitem [{\citenamefont {Maciejko}\ \emph {et~al.}(2009)\citenamefont
  {Maciejko}, \citenamefont {Liu}, \citenamefont {Oreg}, \citenamefont {Qi},
  \citenamefont {Wu},\ and\ \citenamefont {Zhang}}]{halfchelical}%
  \BibitemOpen
  \bibfield  {author} {\bibinfo {author} {\bibfnamefont {J.}~\bibnamefont
  {Maciejko}}, \bibinfo {author} {\bibfnamefont {C.}~\bibnamefont {Liu}},
  \bibinfo {author} {\bibfnamefont {Y.}~\bibnamefont {Oreg}}, \bibinfo {author}
  {\bibfnamefont {X.-L.}\ \bibnamefont {Qi}}, \bibinfo {author} {\bibfnamefont
  {C.}~\bibnamefont {Wu}}, \ and\ \bibinfo {author} {\bibfnamefont {S.-C.}\
  \bibnamefont {Zhang}},\ }\href {\doibase 10.1103/PhysRevLett.102.256803}
  {\bibfield  {journal} {\bibinfo  {journal} {Phys. Rev. Lett.}\ }\textbf
  {\bibinfo {volume} {102}},\ \bibinfo {pages} {256803} (\bibinfo {year}
  {2009})}\BibitemShut {NoStop}%
\bibitem [{\citenamefont {V\"ayrynen}\ and\ \citenamefont
  {Ojanen}(2011)}]{helical1}%
  \BibitemOpen
  \bibfield  {author} {\bibinfo {author} {\bibfnamefont {J.~I.}\ \bibnamefont
  {V\"ayrynen}}\ and\ \bibinfo {author} {\bibfnamefont {T.}~\bibnamefont
  {Ojanen}},\ }\href {\doibase 10.1103/PhysRevLett.107.166804} {\bibfield
  {journal} {\bibinfo  {journal} {Phys. Rev. Lett.}\ }\textbf {\bibinfo
  {volume} {107}},\ \bibinfo {pages} {166804} (\bibinfo {year}
  {2011})}\BibitemShut {NoStop}%
\bibitem [{\citenamefont {Haldane}(1988)}]{Haldane}%
  \BibitemOpen
  \bibfield  {author} {\bibinfo {author} {\bibfnamefont {F.~D.~M.}\
  \bibnamefont {Haldane}},\ }\href {\doibase 10.1103/PhysRevLett.61.2015}
  {\bibfield  {journal} {\bibinfo  {journal} {Phys. Rev. Lett.}\ }\textbf
  {\bibinfo {volume} {61}},\ \bibinfo {pages} {2015} (\bibinfo {year}
  {1988})}\BibitemShut {NoStop}%
\bibitem [{\citenamefont {Nikitov}\ \emph {et~al.}(2001)\citenamefont
  {Nikitov}, \citenamefont {Tailhades},\ and\ \citenamefont {Tsai}}]{magnonic}%
  \BibitemOpen
  \bibfield  {author} {\bibinfo {author} {\bibfnamefont {S.~A.}\ \bibnamefont
  {Nikitov}}, \bibinfo {author} {\bibfnamefont {P.}~\bibnamefont {Tailhades}},
  \ and\ \bibinfo {author} {\bibfnamefont {C.~S.}\ \bibnamefont {Tsai}},\
  }\href@noop {} {\bibfield  {journal} {\bibinfo  {journal} {J. Magn. Magn.
  Mater.}\ }\textbf {\bibinfo {volume} {236}},\ \bibinfo {pages} {320}
  (\bibinfo {year} {2001})}\BibitemShut {NoStop}%
\bibitem [{\citenamefont {Cheuk}\ \emph {et~al.}(2012)\citenamefont {Cheuk},
  \citenamefont {Sommer}, \citenamefont {Hadzibabic}, \citenamefont {Yefsah},
  \citenamefont {Bakr},\ and\ \citenamefont {Zwierlein}}]{MIT}%
  \BibitemOpen
  \bibfield  {author} {\bibinfo {author} {\bibfnamefont {L.~W.}\ \bibnamefont
  {Cheuk}}, \bibinfo {author} {\bibfnamefont {A.~T.}\ \bibnamefont {Sommer}},
  \bibinfo {author} {\bibfnamefont {Z.}~\bibnamefont {Hadzibabic}}, \bibinfo
  {author} {\bibfnamefont {T.}~\bibnamefont {Yefsah}}, \bibinfo {author}
  {\bibfnamefont {W.~S.}\ \bibnamefont {Bakr}}, \ and\ \bibinfo {author}
  {\bibfnamefont {M.~W.}\ \bibnamefont {Zwierlein}},\ }\href {\doibase
  10.1103/PhysRevLett.109.095302} {\bibfield  {journal} {\bibinfo  {journal}
  {Phys. Rev. Lett.}\ }\textbf {\bibinfo {volume} {109}},\ \bibinfo {pages}
  {095302} (\bibinfo {year} {2012})}\BibitemShut {NoStop}%
\bibitem [{\citenamefont {Wang}\ \emph {et~al.}(2012)\citenamefont {Wang},
  \citenamefont {Yu}, \citenamefont {Fu}, \citenamefont {Miao}, \citenamefont
  {Huang}, \citenamefont {Chai}, \citenamefont {Zhai},\ and\ \citenamefont
  {Zhang}}]{ZhaiHui}%
  \BibitemOpen
  \bibfield  {author} {\bibinfo {author} {\bibfnamefont {P.}~\bibnamefont
  {Wang}}, \bibinfo {author} {\bibfnamefont {Z.-Q.}\ \bibnamefont {Yu}},
  \bibinfo {author} {\bibfnamefont {Z.}~\bibnamefont {Fu}}, \bibinfo {author}
  {\bibfnamefont {J.}~\bibnamefont {Miao}}, \bibinfo {author} {\bibfnamefont
  {L.}~\bibnamefont {Huang}}, \bibinfo {author} {\bibfnamefont
  {S.}~\bibnamefont {Chai}}, \bibinfo {author} {\bibfnamefont {H.}~\bibnamefont
  {Zhai}}, \ and\ \bibinfo {author} {\bibfnamefont {J.}~\bibnamefont {Zhang}},\
  }\href {\doibase 10.1103/PhysRevLett.109.095301} {\bibfield  {journal}
  {\bibinfo  {journal} {Phys. Rev. Lett.}\ }\textbf {\bibinfo {volume} {109}},\
  \bibinfo {pages} {095301} (\bibinfo {year} {2012})}\BibitemShut {NoStop}%
\bibitem [{\citenamefont {Lin}\ \emph {et~al.}(2009)\citenamefont {Lin},
  \citenamefont {Compton}, \citenamefont {Perry}, \citenamefont {Phillips},
  \citenamefont {Porto},\ and\ \citenamefont {Spielman}}]{olsf}%
  \BibitemOpen
  \bibfield  {author} {\bibinfo {author} {\bibfnamefont {Y.-J.}\ \bibnamefont
  {Lin}}, \bibinfo {author} {\bibfnamefont {R.~L.}\ \bibnamefont {Compton}},
  \bibinfo {author} {\bibfnamefont {A.~R.}\ \bibnamefont {Perry}}, \bibinfo
  {author} {\bibfnamefont {W.~D.}\ \bibnamefont {Phillips}}, \bibinfo {author}
  {\bibfnamefont {J.~V.}\ \bibnamefont {Porto}}, \ and\ \bibinfo {author}
  {\bibfnamefont {I.~B.}\ \bibnamefont {Spielman}},\ }\href {\doibase
  10.1103/PhysRevLett.102.130401} {\bibfield  {journal} {\bibinfo  {journal}
  {Phys. Rev. Lett.}\ }\textbf {\bibinfo {volume} {102}},\ \bibinfo {pages}
  {130401} (\bibinfo {year} {2009})}\BibitemShut {NoStop}%
\bibitem [{\citenamefont {Barnard}\ \emph {et~al.}(2006)\citenamefont
  {Barnard}, \citenamefont {Xu}, \citenamefont {Li}, \citenamefont {Pradhan},\
  and\ \citenamefont {Peng}}]{CdSe}%
  \BibitemOpen
  \bibfield  {author} {\bibinfo {author} {\bibfnamefont {A.~S.}\ \bibnamefont
  {Barnard}}, \bibinfo {author} {\bibfnamefont {H.}~\bibnamefont {Xu}},
  \bibinfo {author} {\bibfnamefont {X.}~\bibnamefont {Li}}, \bibinfo {author}
  {\bibfnamefont {N.}~\bibnamefont {Pradhan}}, \ and\ \bibinfo {author}
  {\bibfnamefont {X.}~\bibnamefont {Peng}},\ }\href@noop {} {\bibfield
  {journal} {\bibinfo  {journal} {Nanotechnology}\ }\textbf {\bibinfo {volume}
  {17}},\ \bibinfo {pages} {5707} (\bibinfo {year} {2006})}\BibitemShut
  {NoStop}%
\bibitem [{\citenamefont {Xu}\ and\ \citenamefont {Ching}(1993)}]{effectmass}%
  \BibitemOpen
  \bibfield  {author} {\bibinfo {author} {\bibfnamefont {Y.-N.}\ \bibnamefont
  {Xu}}\ and\ \bibinfo {author} {\bibfnamefont {W.~Y.}\ \bibnamefont {Ching}},\
  }\href {\doibase 10.1103/PhysRevB.48.4335} {\bibfield  {journal} {\bibinfo
  {journal} {Phys. Rev. B}\ }\textbf {\bibinfo {volume} {48}},\ \bibinfo
  {pages} {4335} (\bibinfo {year} {1993})}\BibitemShut {NoStop}%
\bibitem [{\citenamefont {Schrier}\ and\ \citenamefont
  {Birgitta~Whaley}(2003)}]{g-factor}%
  \BibitemOpen
  \bibfield  {author} {\bibinfo {author} {\bibfnamefont {J.}~\bibnamefont
  {Schrier}}\ and\ \bibinfo {author} {\bibfnamefont {K.}~\bibnamefont
  {Birgitta~Whaley}},\ }\href {\doibase 10.1103/PhysRevB.67.235301} {\bibfield
  {journal} {\bibinfo  {journal} {Phys. Rev. B}\ }\textbf {\bibinfo {volume}
  {67}},\ \bibinfo {pages} {235301} (\bibinfo {year} {2003})}\BibitemShut
  {NoStop}%
\bibitem [{\citenamefont {Lyssenko}\ \emph {et~al.}(1997)\citenamefont
  {Lyssenko}, \citenamefont {Valu\ifmmode~\check{s}\else \v{s}\fi{}is},
  \citenamefont {L\"oser}, \citenamefont {Hasche}, \citenamefont {Leo},
  \citenamefont {Dignam},\ and\ \citenamefont {K\"ohler}}]{BOsemi1}%
  \BibitemOpen
  \bibfield  {author} {\bibinfo {author} {\bibfnamefont {V.~G.}\ \bibnamefont
  {Lyssenko}}, \bibinfo {author} {\bibfnamefont {G.}~\bibnamefont
  {Valu\ifmmode~\check{s}\else \v{s}\fi{}is}}, \bibinfo {author} {\bibfnamefont
  {F.}~\bibnamefont {L\"oser}}, \bibinfo {author} {\bibfnamefont
  {T.}~\bibnamefont {Hasche}}, \bibinfo {author} {\bibfnamefont
  {K.}~\bibnamefont {Leo}}, \bibinfo {author} {\bibfnamefont {M.~M.}\
  \bibnamefont {Dignam}}, \ and\ \bibinfo {author} {\bibfnamefont
  {K.}~\bibnamefont {K\"ohler}},\ }\href {\doibase 10.1103/PhysRevLett.79.301}
  {\bibfield  {journal} {\bibinfo  {journal} {Phys. Rev. Lett.}\ }\textbf
  {\bibinfo {volume} {79}},\ \bibinfo {pages} {301} (\bibinfo {year}
  {1997})}\BibitemShut {NoStop}%
\bibitem [{\citenamefont {Ben~Dahan}\ \emph {et~al.}(1996)\citenamefont
  {Ben~Dahan}, \citenamefont {Peik}, \citenamefont {Reichel}, \citenamefont
  {Castin},\ and\ \citenamefont {Salomon}}]{BOoptical}%
  \BibitemOpen
  \bibfield  {author} {\bibinfo {author} {\bibfnamefont {M.}~\bibnamefont
  {Ben~Dahan}}, \bibinfo {author} {\bibfnamefont {E.}~\bibnamefont {Peik}},
  \bibinfo {author} {\bibfnamefont {J.}~\bibnamefont {Reichel}}, \bibinfo
  {author} {\bibfnamefont {Y.}~\bibnamefont {Castin}}, \ and\ \bibinfo {author}
  {\bibfnamefont {C.}~\bibnamefont {Salomon}},\ }\href {\doibase
  10.1103/PhysRevLett.76.4508} {\bibfield  {journal} {\bibinfo  {journal}
  {Phys. Rev. Lett.}\ }\textbf {\bibinfo {volume} {76}},\ \bibinfo {pages}
  {4508} (\bibinfo {year} {1996})}\BibitemShut {NoStop}%
\bibitem [{\citenamefont {Zhu}\ \emph {et~al.}(2009)\citenamefont {Zhu},
  \citenamefont {Zhang},\ and\ \citenamefont {Wang}}]{1DS2}%
  \BibitemOpen
  \bibfield  {author} {\bibinfo {author} {\bibfnamefont {S.-L.}\ \bibnamefont
  {Zhu}}, \bibinfo {author} {\bibfnamefont {D.-W.}\ \bibnamefont {Zhang}}, \
  and\ \bibinfo {author} {\bibfnamefont {Z.~D.}\ \bibnamefont {Wang}},\ }\href
  {\doibase 10.1103/PhysRevLett.102.210403} {\bibfield  {journal} {\bibinfo
  {journal} {Phys. Rev. Lett.}\ }\textbf {\bibinfo {volume} {102}},\ \bibinfo
  {pages} {210403} (\bibinfo {year} {2009})}\BibitemShut {NoStop}%
\bibitem [{\citenamefont {B\"uttiker}\ \emph {et~al.}(1985)\citenamefont
  {B\"uttiker}, \citenamefont {Imry}, \citenamefont {Landauer},\ and\
  \citenamefont {Pinhas}}]{LB}%
  \BibitemOpen
  \bibfield  {author} {\bibinfo {author} {\bibfnamefont {M.}~\bibnamefont
  {B\"uttiker}}, \bibinfo {author} {\bibfnamefont {Y.}~\bibnamefont {Imry}},
  \bibinfo {author} {\bibfnamefont {R.}~\bibnamefont {Landauer}}, \ and\
  \bibinfo {author} {\bibfnamefont {S.}~\bibnamefont {Pinhas}},\ }\href
  {\doibase 10.1103/PhysRevB.31.6207} {\bibfield  {journal} {\bibinfo
  {journal} {Phys. Rev. B}\ }\textbf {\bibinfo {volume} {31}},\ \bibinfo
  {pages} {6207} (\bibinfo {year} {1985})}\BibitemShut {NoStop}%
\bibitem [{\citenamefont {Mahan}(2000)}]{GF}%
  \BibitemOpen
  \bibfield  {author} {\bibinfo {author} {\bibfnamefont {G.~D.}\ \bibnamefont
  {Mahan}},\ }\href@noop {} {\emph {\bibinfo {title} {Many-Particle
  Physics}}},\ \bibinfo {edition} {3rd}\ ed.\ (\bibinfo  {publisher} {Plenum},\
  \bibinfo {address} {New York},\ \bibinfo {year} {2000})\BibitemShut {NoStop}%
\bibitem [{\citenamefont {Anderson}\ \emph {et~al.}(1980)\citenamefont
  {Anderson}, \citenamefont {Thouless}, \citenamefont {Abrahams},\ and\
  \citenamefont {Fisher}}]{1Dlocal1}%
  \BibitemOpen
  \bibfield  {author} {\bibinfo {author} {\bibfnamefont {P.~W.}\ \bibnamefont
  {Anderson}}, \bibinfo {author} {\bibfnamefont {D.~J.}\ \bibnamefont
  {Thouless}}, \bibinfo {author} {\bibfnamefont {E.}~\bibnamefont {Abrahams}},
  \ and\ \bibinfo {author} {\bibfnamefont {D.~S.}\ \bibnamefont {Fisher}},\
  }\href {\doibase 10.1103/PhysRevB.22.3519} {\bibfield  {journal} {\bibinfo
  {journal} {Phys. Rev. B}\ }\textbf {\bibinfo {volume} {22}},\ \bibinfo
  {pages} {3519} (\bibinfo {year} {1980})}\BibitemShut {NoStop}%
\bibitem [{\citenamefont {Jackiw}\ and\ \citenamefont {Rebbi}(1976)}]{half1}%
  \BibitemOpen
  \bibfield  {author} {\bibinfo {author} {\bibfnamefont {R.}~\bibnamefont
  {Jackiw}}\ and\ \bibinfo {author} {\bibfnamefont {C.}~\bibnamefont {Rebbi}},\
  }\href {\doibase 10.1103/PhysRevD.13.3398} {\bibfield  {journal} {\bibinfo
  {journal} {Phys. Rev. D}\ }\textbf {\bibinfo {volume} {13}},\ \bibinfo
  {pages} {3398} (\bibinfo {year} {1976})}\BibitemShut {NoStop}%
\end{thebibliography}%

\end{document}